\documentclass[pdflatex,sn-mathphys]{sn-jnl}

\theoremstyle{thmstyleone}%
\theoremstyle{thmstyletwo}%

\theoremstyle{thmstylethree}%

\usepackage{xr-hyper} 
\makeatletter

\usepackage{graphicx}
\usepackage{dcolumn}
\usepackage{bm}

\usepackage{xcolor}
\colorlet{rev0}{black}

\usepackage{breakurl}
\usepackage{url}

\usepackage{amsmath}
\usepackage{xcite}
\usepackage{siunitx} 
 
\usepackage[export]{adjustbox} 
\usepackage[percent]{overpic}

\usepackage{hyperref}

\begin{document}



\title{Towards Universal Urban Patterns-of-Life Simulation}


\author*[1]{\fnm{Sandro Martinelli} \sur{Reia}}\email{smarti71@gmu.edu}

\author[1]{\fnm{Henrique Ferraz} \sur{de Arruda}}

\author[1]{\fnm{Shiyang} \sur{Ruan}}

\author[1]{\fnm{Taylor} \sur{Anderson}}

\author[2]{\fnm{Hamdi} \sur{Kavak}}

\author[1]{\fnm{Dieter} \sur{Pfoser}}

\affil[1]{
\orgdiv{Geography and Geoinformation Science, College of Science}, 
\orgname{George Mason University},
\orgaddress{\street{4400 University Dr.}, 
\city{Fairfax}, 
\postcode{22030}, 
\state{Virginia}, 
\country{USA}}}

\affil[2]{
\orgdiv{Center for Social Complexity, College of Science},
\orgname{George Mason University},
\orgaddress{\street{ 4400 University Dr.}, 
\city{Fairfax}, 
\postcode{22030}, 
\state{Virginia}, 
\country{USA}}}


\abstract{
Understanding urban mobility requires models that capture how people interact with and navigate the built environment.
We present a scalable, generalizable agent-based framework in which daily schedules emerge from the interplay between mandatory (e.g., work, school) and flexible (e.g., errands, food, leisure) activities, driven by evolving individual needs.
The results of our model are validated against empirical patterns from the 2017 U.S. National Household Travel Survey, including activity distributions, origin–destination flows, and trip-chain length distributions.
We introduce a normalized similarity metric to quantify agreement between simulated and empirical patterns.
Most cities achieve scores above 0.80, demonstrating strong alignment without the need for city-specific calibration.
The model scales efficiently to over 20 million agents, enabling full-population simulations of large metropolitan areas.
This combination of universality and scalability enables scenario analysis for infrastructure stress testing, disaster recovery, innovation diffusion, and disease spread in urban systems.
}


\maketitle
\newpage 
\section{Introduction}\label{sec1}

Simulating human mobility has important applications across urban planning, transportation systems, epidemiology, and disaster response. Accurate representations of how individuals move within and across regions enable the design of more efficient infrastructure, forecasting congestion and disease spread, and the evaluation of policy interventions, aligning with the goals of sustainable and resilient cities~\cite{UN_SDGs}. 

At the individual level, a key challenge in mobility simulation is determining the sequence of activities (trip chains) people carry out over a typical day. Empirical studies have shown that these chains often follow a compact and regular structure of a small number of motifs, typically beginning and ending at home that account for the majority of daily routines~\cite{primerano2008defining,schneider2013unravelling}. These motifs can be inferred from mobile-phone data, GPS traces, or traditional travel surveys~\cite{alexander2015origin,wang2018applying,holguin2005observed,mcguckin2004trips}. Such patterns exhibit regularity and diversity, aligning with findings from statistical physics and complexity science that demonstrate how human movement tends to follow reproducible and predictable patterns~\cite{Gonzalez2008Nature,Barbosa2018PhysRep}.

Simulation strategies to reproduce such chains often follow a top-down approach, in which activity schedules are generated from probabilistic rules, utility-based formulations, or generative models trained on empirical data~\cite{felbermair2020generating,sallard2021open,adnan2016simmobility}. In these frameworks, daily plans are either synthesized from statistical distributions or directly sampled from large-scale travel surveys and mobility traces. Utility-maximizing approaches such as CEMDAP~\cite{bhat2004comprehensive} construct schedules by chaining discrete choices about activity type, timing, and location, while newer probabilistic models leverage sequence trees, Hidden Markov Models, or variational autoencoders to synthesize chains that match empirical distributions~\cite{Drchal2019TRC,Ouyang2018IJCAI,Huang2019MIPR,Yin2018TITS}. These models can reproduce observed behavior with high fidelity, but often rely on fixed rules, extensive calibration to local datasets, or access to large amounts of region-specific mobility traces, which limits their transferability and raises privacy concerns~\cite{Hess2015ACMCS}.

In contrast, agent-based models (ABMs) adopt a bottom-up view in which autonomous agents with heterogeneous preferences, constraints, and bounded rationality interact with a structured environment, producing mobility as an emergent outcome~\cite{crooks2008key,bonabeau2002agent,folsom2013scalable,reia2019agent,pflieger2010introduction,bassolas2019hierarchical}. ABMs offer greater behavioral expressiveness and are capable of simulating individual decision-making in dynamic environments, allowing for policy-sensitive what-if analyses and exploration of complex adaptive behaviors~\cite{Rasouli2014ABMReview}. 
ABMs have been used to model everything from mode choice and activity scheduling~\cite{arentze2000albatross, adnan2016simmobility} to urban segregation~\cite{yin2009dynamics} and infrastructure recovery~\cite{moradi2020recovus, xue2023supporting}, often operating at high spatial and temporal resolution. 
However, this expressive power comes at the cost of increased computational burden, and model outcomes are sensitive to the choice of rules, parameters, and initialization~\cite{bonabeau2002agent,auld2016polaris}.


Despite progress, some key gaps remain. 
First, many large-scale mobility models are tightly tuned to specific locations and struggle to generalize~\cite{auld2016polaris,adnan2016simmobility}, thus limiting their portability and introducing a reliance on detailed, often hard-to-acquire datasets. 
Second, validation practices are often not robust. Limits to generalization mean that few models are tested in multiple cities or against multiple empirical targets, and even fewer control for population composition differences between synthetic population data and observed datasets~\cite{grimm2005pattern,NHTS2017,USCensus2020DP1,USCensusACS2017S2301}. 
These issues are especially salient when using datasets that are not representative of certain population groups. Third, scalability remains a persistent limitation. 
Behaviorally rich ABMs are computationally intensive and therefore are rarely implemented on scales sufficient to model entire metropolitan populations, restricting their utility for real-world decision making~\cite{bonabeau2002agent}.

This paper presents a modeling framework that addresses these gaps by showing that human mobility patterns can be reproduced across multiple U.S. metropolitan areas using a single, generalizable model with a standard set of behavioral parameters. 
Our approach combines first-principles representations of behavioral drivers such as mandatory activities such as work and school, and flexible, needs-driven activities such as food, errands, and recreation. The population composition is derived from Census data and infrastructure extracted from OpenStreetMap~\cite{OpenStreetMap,USCensus2020DP1,USCensusACS2017S2301}. 
Rather than prescribing activity chains, which are given by the sequence of places visited in a 24 hour period~\cite{mcguckin2004trips}, the model allows schedules to emerge from evolving needs and spatial constraints. 
Destination choice follows empirical concentration patterns using rank-based probabilities~\cite{song2010modelling}.

To evaluate model performance, we adopt a pattern-oriented modeling approach~\cite{grimm2005pattern} and validate simultaneously against multiple empirical patterns from the 2017 U.S. National Household Travel Survey~\cite{NHTS2017}. These include the frequency of trips to each activity category, origin–destination flow structures, and the distribution of number of trips.
Importantly, we account for differences in population composition between the NHTS and the synthetic populations derived from Census microdata, ensuring that comparisons are not biased by sampling effects~\cite{USCensus2020DP1,USCensusACS2017S2301}. 
Results show that a single behavioral model parameterization can reproduce the dominant mobility patterns in most cities studied, with an overall similarity score above $80\%$.
In cases where infrastructure imposes longer travel distances, small and interpretable adjustments (like increasing effective travel speed) recover expected activity patterns without modifying the behavioral core of the model.
The model can be scaled to tens of millions of agents with complete daily schedules, supporting city-scale analyses in mobility, resilience, and equity~\cite{collier2022distributed}. 


In what follows, we analyze results from cross-city comparisons (Section~\ref{secResults}), discuss the broader implications for understanding, forecasting, and planning urban mobility systems (Section~\ref{secDiscussion}), and present a description of the model’s architecture and input data in detail (Section~\ref{secMethods}).

\section{Results}
\label{secResults}

\begin{figure}
    \centering
    \includegraphics[width=1\linewidth]{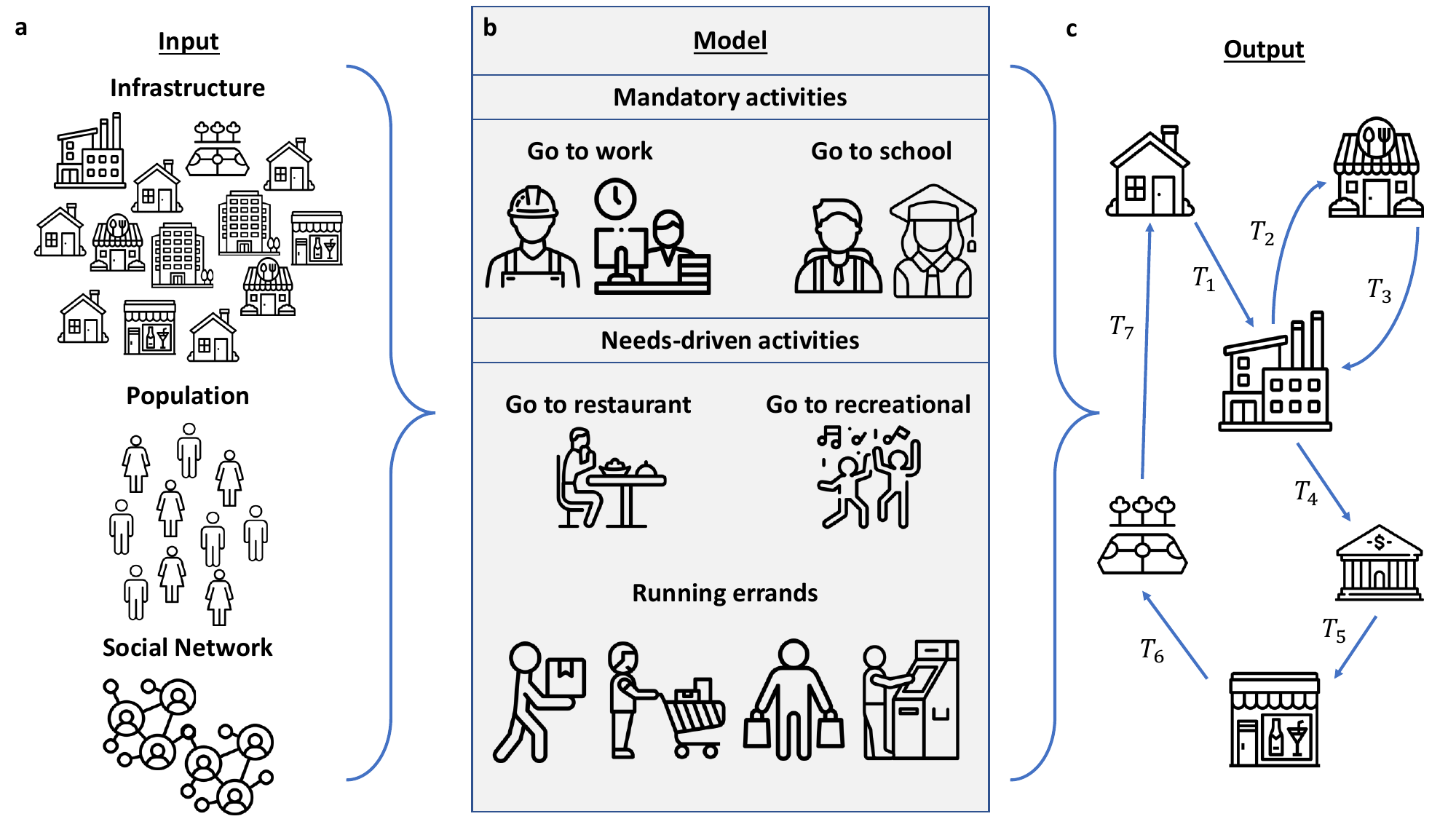}
    \includegraphics[width=1\linewidth]{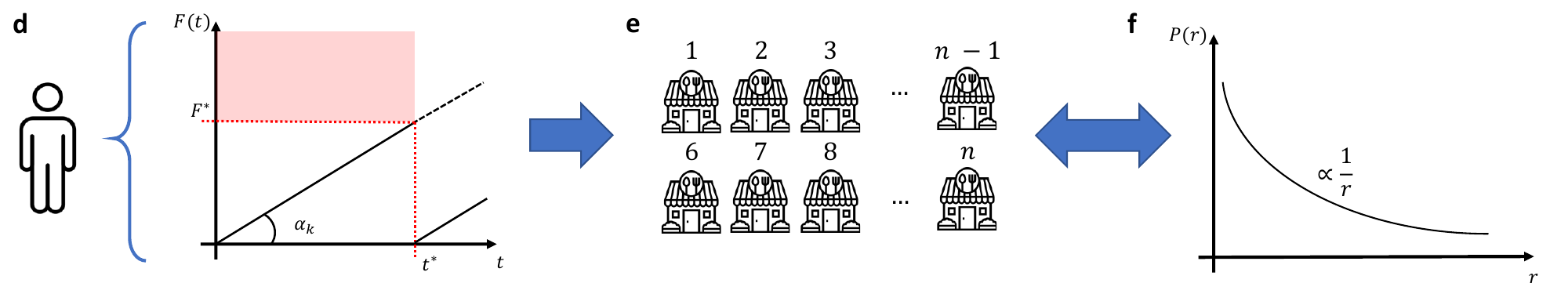}
    \caption{\textbf{Conceptual diagram of the agent-based model proposed here.}
    The model’s input (a) consists of the spatial distribution of infrastructure, population composition, sociodemographic attributes, and social networks. Infrastructure includes residential buildings, where agents live and spend the night, and non-residential buildings, where agents go to work, school, restaurants, recreational places, and run errands. 
    The model (b) assigns both mandatory and needs-driven activities to agents. Mandatory activities depend on agent type: workers, students, and homemakers. Workers must go to work, students attend school, and homemakers do not engage in either work or school activities.
    The model’s output (c) is a sequence of trips starting and ending at home, representing agents’ daily patterns of life. In this example, the agent leaves home for work ($T_1$), goes to a restaurant for lunch ($T_2$), returns to work ($T_3$), then after work runs an errand ($T_4$), goes to a restaurant ($T_5$), visits a recreational place ($T_6$), and finally returns home ($T_7$).
    Each agent is characterized by three parameters that represent the intensity of growth of their food, social, and errand needs (d). When a need intensity exceeds a threshold (set here at $1$), a trip is triggered to a building associated with that activity. Buildings are chosen from a set (e) according to a probability distribution (f) that reflects the agent’s preference for each building. In other words, each agent is assigned a predefined list of potential destinations together with a ranked preference ordering.
    By combining these components, the model generates detailed activity schedules for all agents, reflecting realistic patterns of life.}
    \label{fig_model}
\end{figure}

Population and infrastructure serve as inputs to the agent-based model we propose here (\autoref{fig_model}).
The three types of agents in our simulation, which are workers, homemakers and students, engage in both mandatory and needs-driven activities.
Workers and students have work and school as their respective mandatory activities, while homemakers are agents who do not participate in work or school.
Needs-driven activities are triggered by agents’ internal states, given by their needs for food, social interaction, and errands (see Section \ref{secMethods}), which represent broader, non-explicit activity categories in the NHTS.
When a need exceeds a given threshold, the agent initiates a trip to a respective activity location. 
The choice of destination is determined by the agent’s preference for each building, with the visitation probability following a power-law decay $P(r) \propto r^{-\xi}$, where $r$ denotes the $r$-th most preferred location~\cite{song2010modelling}.
We have run the simulation for $7$ cities with population higher than $1M$, which are Minneapolis, Riverside, Hartford, Milwaukee, San Jose, Cleveland and New York.
The simulation, which outputs the sequence of trips for each agent, is run for two week days, capturing the typical behavior of agents in a regular day.

\begin{figure}[t]
    \includegraphics[width=0.895\linewidth]{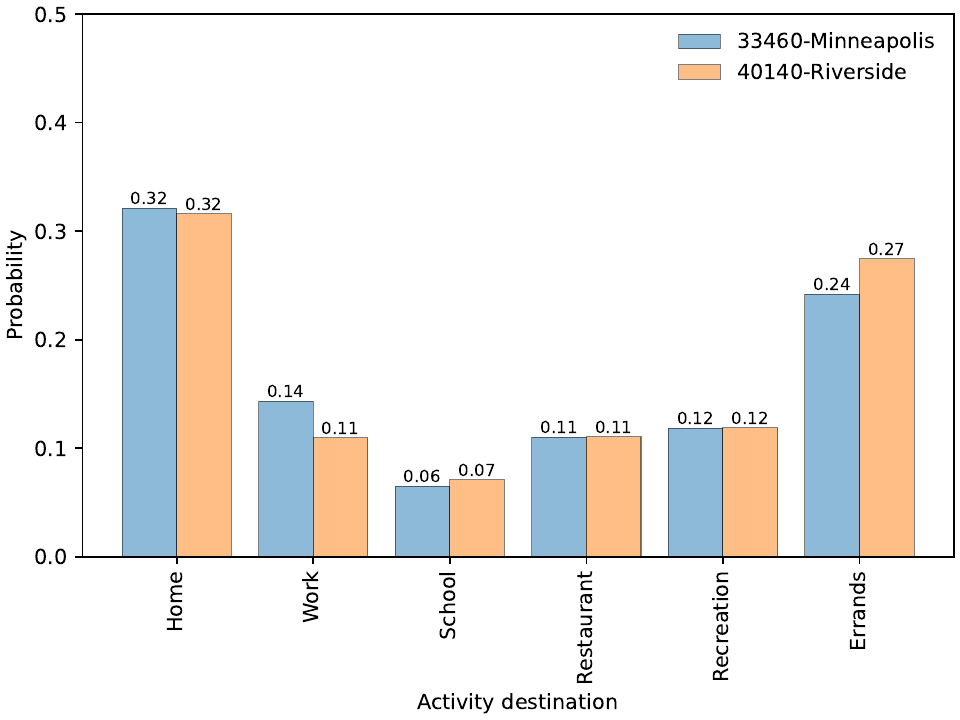}
    \caption{\textbf{Same set of parameters may lead to different activity patterns across cities.} 
    Simulation results showing the probability distribution of activity destinations, home, work, school, restaurant, recreation, and errands for two metropolitan areas, Minneapolis Riverside, using the same set of model parameters.
    Despite identical behavioral rules and parameter values in the agent-based model, the resulting activity patterns differ across cities, driven by differences in population composition (e.g., higher proportions of workers in the Minneapolis, and higher proportion of homemakers in the Riverside).
    }
    \label{fig_freq_cat_High_Low_Sim}
\end{figure}

\subsection{Drivers of patterns of life}

The empirical population composition extracted from the NHTS (Supplementary Table 1) shows that in most cities, workers account for about $50\%$ of the population, with the remaining population composed of students (about $15\%$) and homemakers (about $35\%$).
We observe that the frequency of trips for mandatory activities is determined by the types of agents.
A high proportion of workers leads to more trips to work, while a high proportion of homemakers leads to more errand-related trips (see Supplementary Fig. 2).
We can also see that cities with similar overall population compositions, such as Austin and San Jose, tend to have similar patterns, with similar visitation frequencies to the six categories explored here.
Since our study focuses on metropolitan areas, we use the terms ``city'' and ``metropolitan area'' interchangeably, referring to them by the first name of the metropolitan area, e.g., New York for the New York-Newark-Jersey City, NY-NJ-PA metropolitan area.
The analysis of the population composition and resulting fraction of trips for all cities reported in the NHTS reveals correlations between the proportion of workers and the fraction of trips to workplaces, as well as the proportion of students and homemakers and the fraction of trips to schools and errand places, respectively (see Supplementary Fig. 3).

The effect of population composition on patterns of life is also observed in our model.  
While trips to mandatory activities are determined by the agents' type, trips to needs-driven activities are triggered by the agents' need intensity, which is an adjustable parameter of the simulation (see \ref{secMethods} section).  
Consequently, the same set of parameters can lead to different outcomes in different cities, as observed when comparing the model results for Minneapolis and Riverside (see \autoref{fig_freq_cat_High_Low_Sim}).  
Consider the $3.7$ million population of Minneapolis (see Supplementary Fig. 5), which is composed of $53\%$ workers, $24\%$ students, and $23\%$ homemakers and the $4.6$ million population of Riverside (see Supplementary Fig. 6) consists of $42\%$ workers, $27\%$ students, and $32\%$ homemakers.  
This difference in population composition is reflected in a higher number of trips to workplaces in Minneapolis (higher percentage of workers) and more errands-related trips in Riverside (higher percentage of homemakers).

\begin{figure}
    \centering
    \includegraphics[width=0.875\linewidth]{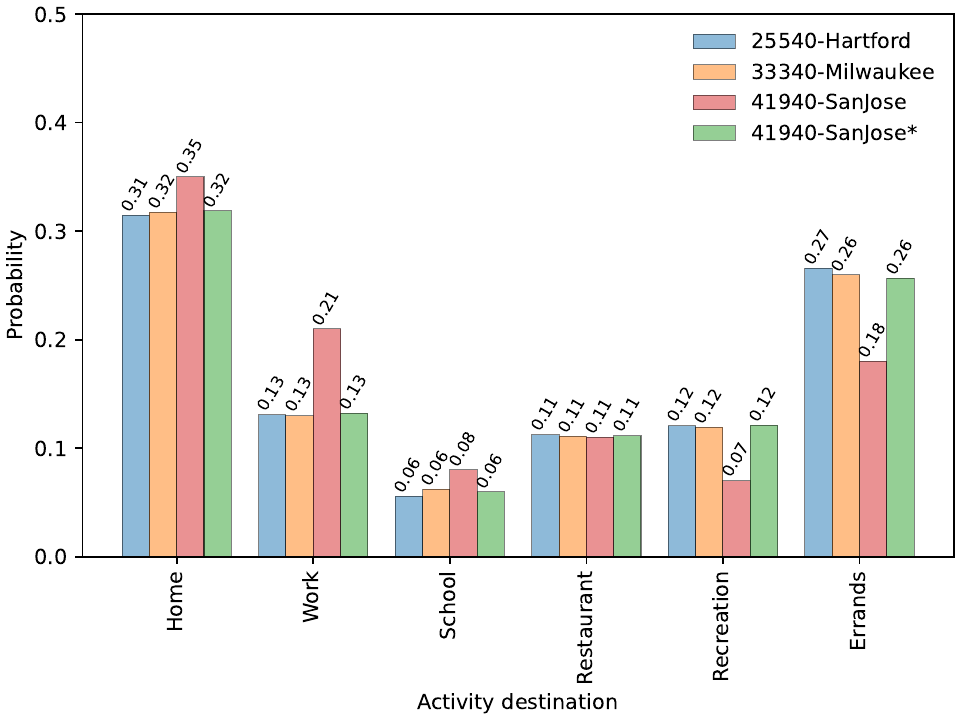}
    \caption{\textbf{Patterns of life are constrained by infrastructure configuration.}
    The cities shown here, Hartford, Milwaukee, and San Jose, have similar population compositions in terms of the proportions of workers, students, and homemakers. This similarity was expected to produce comparable simulation results under the same set of parameters. However, the spatial layout of San Jose leads to longer home-to-work distances, which limits the number of trips agents can complete in a day and reduces the frequency of flexible activities such as recreation and errands. When a single parameter (travel speed) is adjusted, agents in San Jose are able to fit more trips into their daily schedules, thereby reproducing the expected activity frequency distribution (indicated by *).
    }
    \label{fig_sim_infrastructure}
\end{figure}

In cities with similar population composition, the same set of parameters leads to similar patterns of life, as observed for Hartford, Milwaukee and San Jose (\autoref{fig_sim_infrastructure}). 
Hartford (see Supplementary Fig. 7) has a population of approximately $1.2$ million, composed of $50, 21, 29\%$ of workers, students, and homemakers, respectively.  
Milwaukee (see Supplementary Fig. 8), with a population of $1.6$ million, has $49, 23, 28\%$ of workers, students, and homemakers.  
The $2.0$ million population of San Jose (see Supplementary Fig. 9) has a $51, 22, 27\%$ worker, student, homemaker split.
The patterns for Hartford, with an average trip distance of $9.6\,\mathrm{km}$ per day, are similar to those of Milwaukee, where agents travel $9.2\,\mathrm{km}$ per trip per day, as expected.  
However, in our simulation, San Jose has more work trips and fewer recreation and errand trips.  
With an average trip distance of $16.3\,\mathrm{km}$, agents in San Jose spend more time traveling to mandatory activities like school and work, and have less time to fulfill needs-driven activities such as going to recreational places and running errands.

In fact, empirical analysis of NHTS data reveals that populations in cities adjust their travel behavior to accommodate infrastructure constraints in large cities (see Supplementary Fig. 4).  
The average number of trips per person per day decreases with increased travel distance. This means that people traveling longer distances have fewer opportunities to explore the city.  
An immediate consequence in cities where people travel longer distances is increased travel time, with people spending more time in traffic.  
The relationship between travel time and average distance is sublinear (correlation coefficient less than 1), due to for example better developed road networks in larger cities that support on average higher capacities and speeds.
Using this insight, we increase the possible speeds in San Francisco. A reduced time in traffic allows agents to cover longer distances faster, thus removing the distance constraints imposed by infrastructure.  
The revised simulation results exhibit patterns that match those of other cities with similar population composition and needs parameters (\autoref{fig_sim_infrastructure}, San Jose*).

\subsection{Assessing behavioral realism}

NHTS can be used to verify the results generated by our model. To do so we have to first assess the similarities in population composition between our model and NHTS.  
Our model uses Census/ACS to derive the population composition of the study areas (see Supplementary Table 2) and the comparison to NHTS shows a positive correlation between the fractions of workers, students, and homemakers in both populations. However, in NHTS students are underrepresented while workers and homemakers are overrepresented (Supplementary Fig. 12).  
As such, our simulation will generate more school trips and fewer work and errand trips when compared to NHTS given this population imbalance.

A ranking of cities by differences in population composition (see Supplementary Table 3) between Census and NHTS reveals that Cleveland has the smallest difference.  
Cleveland is a metropolitan area with a population of $2.1$ million and a $48, 21,31\%$ worker, student, homemaker split according to Census data. NHTS data captures $382$ respondents, of whom $50\%$ were considered workers, $17\%$ students, and $33\%$ homemakers, which is comparable to the Census data.

\begin{figure}
    \begin{overpic}[width=0.48\linewidth,height=0.28\textheight,keepaspectratio,valign=c]{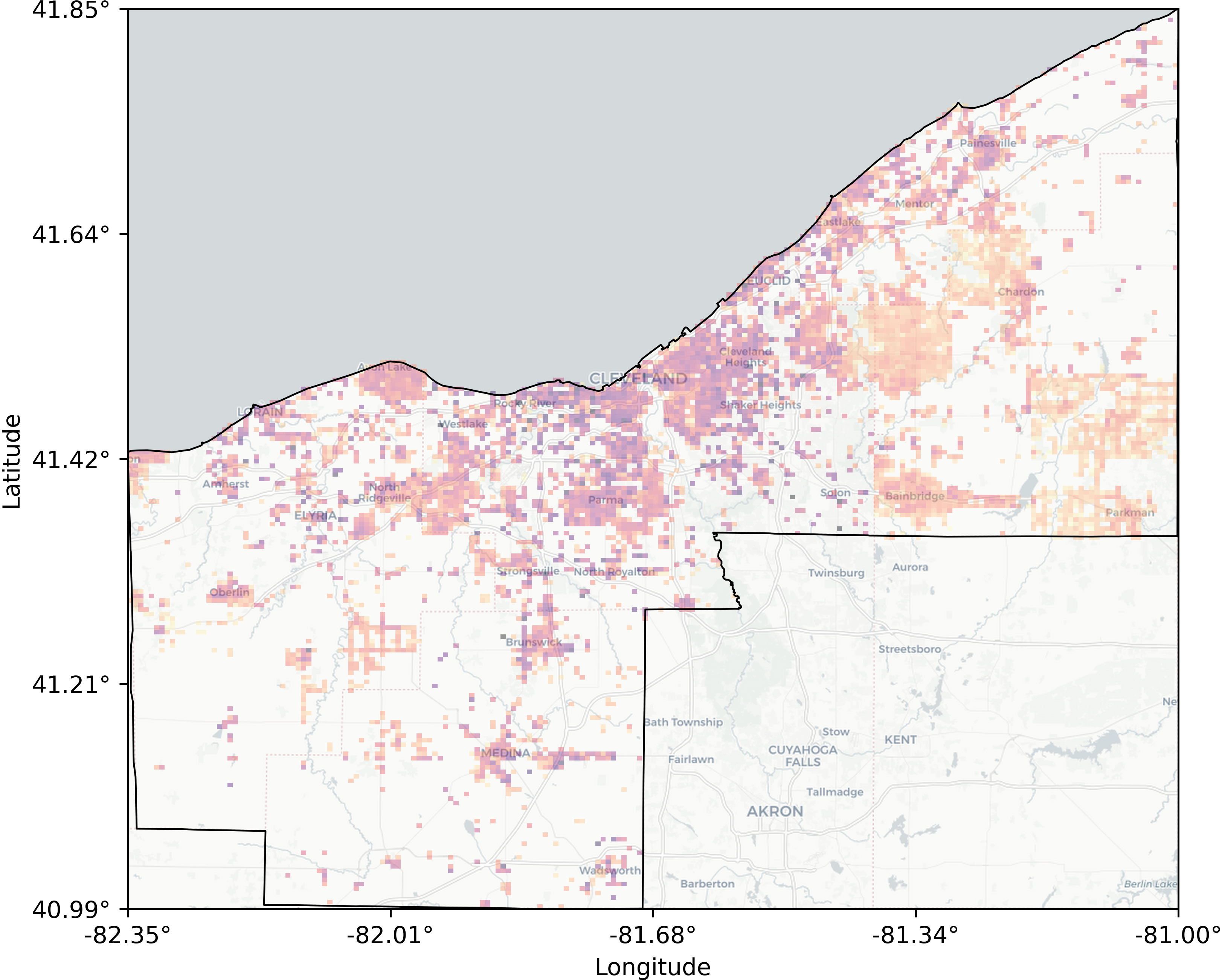}
        \put(2,85){\textbf{a}}
    \end{overpic}%
    \begin{overpic}[width=0.48\linewidth,height=0.28\textheight,keepaspectratio,valign=c]{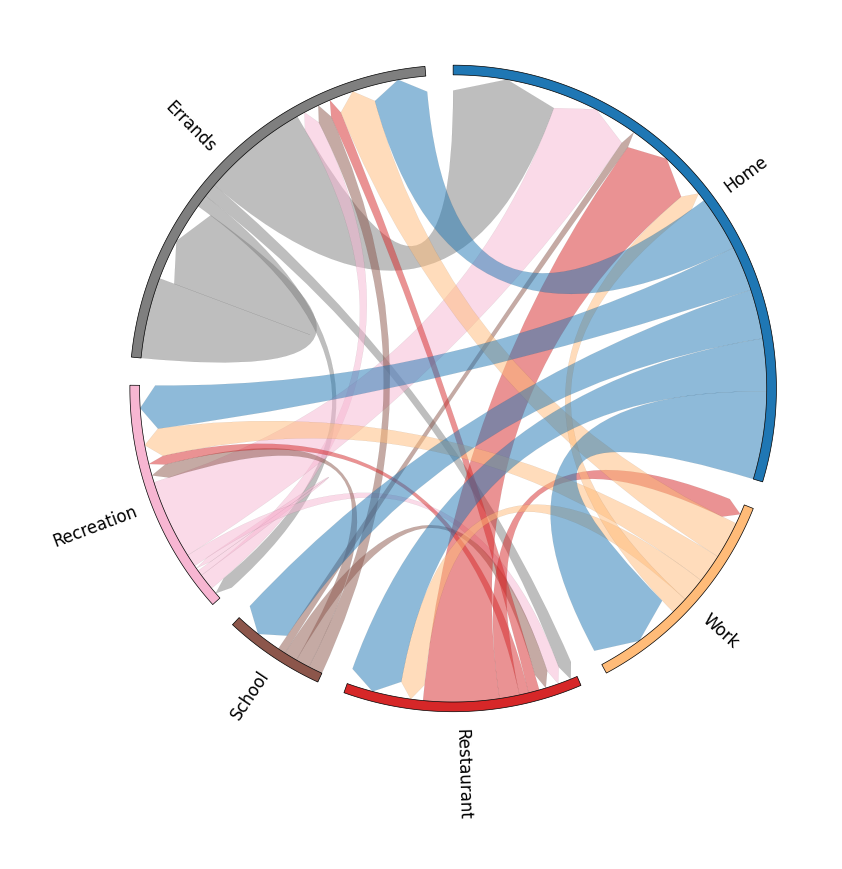}
        \put(2,87.5){\textbf{d}}
    \end{overpic}

    \begin{overpic}[width=0.48\linewidth,height=0.28\textheight,keepaspectratio,valign=c]{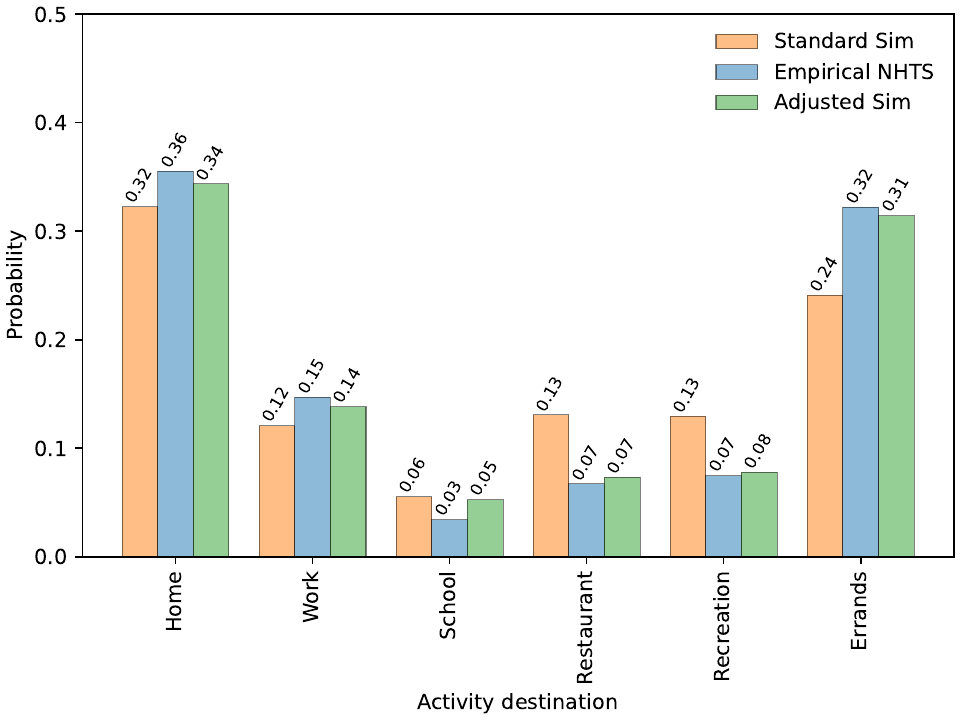}
        \put(2,85){\textbf{b}}
    \end{overpic}%
    \begin{overpic}[width=0.48\linewidth,height=0.28\textheight,keepaspectratio,valign=c]{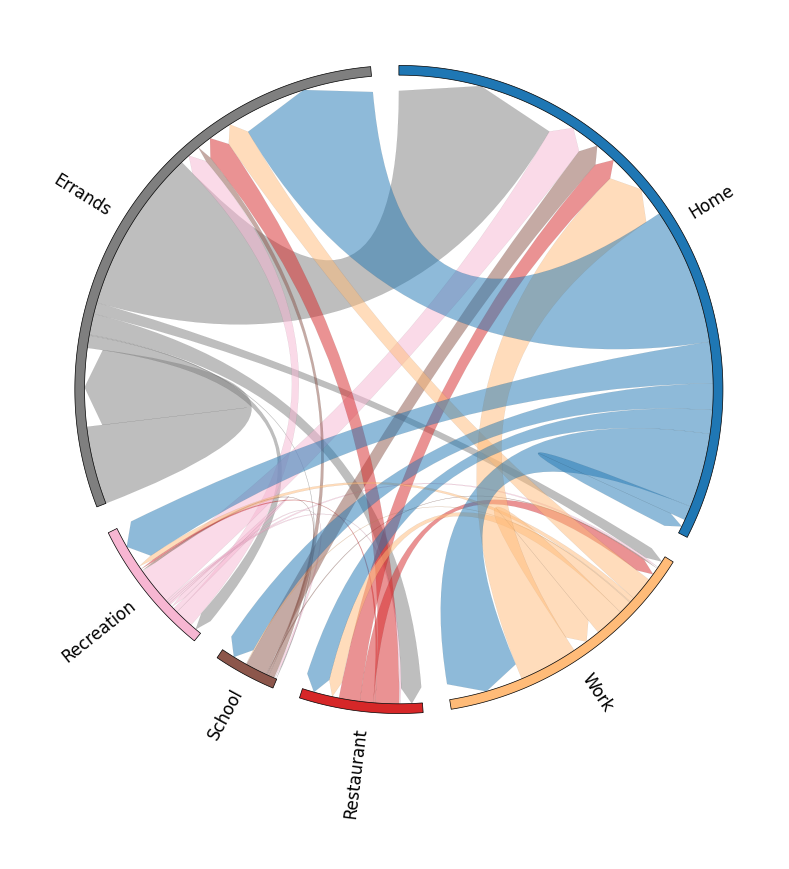}
        \put(2,87.5){\textbf{e}}
    \end{overpic}

    \begin{overpic}[width=0.48\linewidth,height=0.28\textheight,keepaspectratio,valign=c]{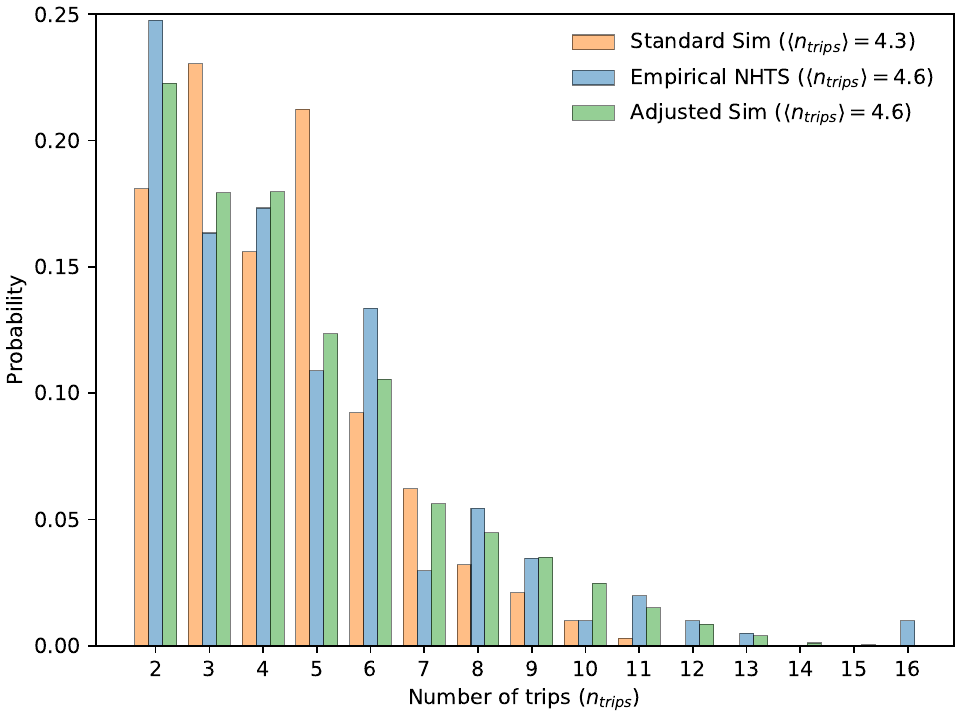}
        \put(2,85){\textbf{c}}
    \end{overpic}%
    \begin{overpic}[width=0.48\linewidth,height=0.28\textheight,keepaspectratio,valign=c]{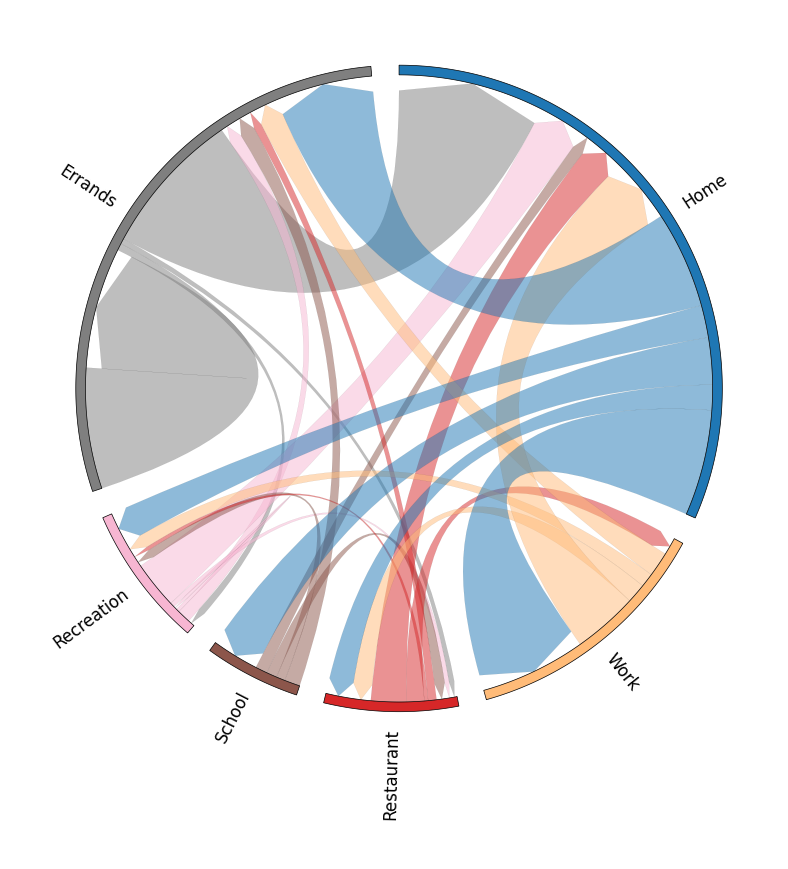}
        \put(2,87.5){\textbf{f}}
    \end{overpic}

    \caption{\textbf{Empirical trends approximated with simulation standard parameters are refined by adjustment.}
    Panel (a) shows the spatial distribution of the population in the Cleveland metropolitan area, where darker colors indicate higher density. Panel (b) presents the frequency of visits to different activity destinations. Results are compared across the standardized simulation, the empirical distributions from the NHTS, and an adjusted version of the simulation. Panel (c) depicts the distribution of the number of daily trips per person, again contrasting standardized, empirical, and adjusted cases. Panels (d)–(f) show trip flow diagrams that capture the movement of agents between activity types. Panel (d) corresponds to the standardized simulation, illustrating baseline patterns. Panel (e) represents empirical flows derived from the NHTS, providing real-world reference. Panel (f) shows the adjusted simulation, which better aligns with empirical observations by calibrating model parameters. Together, these panels illustrate how model calibration improves the agreement between simulated and observed mobility patterns across both activity frequencies and trip flows.}
    \label{fig_sim_NHTS_comparison}
\end{figure}

Cleveland, OH, is adjacent to Lake Erie, with its population concentrated along road networks near the lake and in the central areas of the metropolitan region (see \autoref{fig_sim_NHTS_comparison}, Supplementary Fig. 10).
Using standardized parameters provides a first approximation of patterns of life in Cleveland. 
However, this standard version underestimates Errand trips and overestimates trips to Restaurants and Recreational places when compared to NHTS data.
It also produces shorter trip chains, with an average of $4.3$ trips per person per day. 
While the flow diagram highlights the complexity of intertwined activity chains, it also reveals an overrepresentation of trips from Home to Restaurants and Recreation, and an underrepresentation of Work-related trips (\autoref{fig_sim_NHTS_comparison}).

Based on this mismatch, we tuned the behavioral parameters that control how often agents trigger needs-driven activities (see Supplementary Fig. 13).
For example, the overrepresentation of trips from Home to Restaurants was reduced by decreasing the agent's need for food, while the underestimation of Errand trips was corrected by slightly increasing the needs for running errands.
As a result, the most common destinations, Home, Errands, and Work, are followed by Recreation, Restaurant, and School. 
The fraction of trips to each activity type differs from the NHTS by less than $0.02\%$, with the only noticeable discrepancy being slightly fewer school trips due to the underrepresentation of students in the NHTS population sample. 
The trip flow diagram confirms this closer alignment, showing strong flows between Home and Errands, Home and Work, and Home and School. 
The high frequency of trips to and from Home underscores its role as an activity hub \cite{schneider2013unravelling, reia2025function}, while the prevalence of Errand-to-Errand trips indicates the chaining of these activities into longer sequences.

The adjusted simulation also reproduces the empirical average of $\langle n_{\mathrm{trips}} \rangle = 4.6$ trips per person per day and closely matches the probability distribution of trip chains (\autoref{fig_sim_NHTS_comparison}). 
The most frequent chains consist of two trips, followed by chains of four and three trips. 
This distribution is marked by three significant gaps after $n_{\mathrm{trips}} = 2$, $4$, and $6$. 
The rank plot in Supplementary Fig. 14 captures the disparity and complexity of trip chain generation, revealing a hierarchical structure in their distribution. 
This structure suggests the existence of preferential activity choices, reproducing patterns observed in the literature \cite{reia2025function, ectors2019exploratory}.


\subsection{Universality of patterns of life}

To compare simulated and empirical patterns of life across cities, we represent each pattern as a probability vector $\bold{x} = \left( x_1, \dots, x_n \right)$ such that $\sum_i{x_i} = 1, x_i \geq 0$.
We consider the following patterns of life $p$, 
(i) the activity-destination distribution, 
(ii) the distribution of trip-chain lengths $n_{trips}$, and 
(iii) origin-destination flow diagrams.
For each pattern, we determine the simulation ($\bold{x}^{p}_{sim}$) and the NHTS ($\bold{x}^{p}_{NHTS}$) vectors in the same state space.
By obtaining the Euclidean distance ($d^p = || \bold{x}^{p}_{sim} - \bold{x}^{p}_{NHTS} ||$) and normalizing it by the maximum possible distance on the simplex ($d_{max} = \sqrt{2}$), the pattern-specific similarity is given by:

\begin{equation}
  S^{(p)} \;=\; 1 - \frac{d^{(p)}}{d_{\max}}
  \;=\; 1 - \frac{d^{(p)}}{\sqrt{2}}
  \;\in\; [0,1].
  \label{eq:similarity}
\end{equation}

The similarity scores reported in \autoref{tab:sim-nhts-similarity-2dp} show that the standard model captures the patterns of the cities considered very well, with most cities having at least $80\%$ similarity in the three patterns.
Without specific calibration, Minneapolis has the best agreement with the empirical data: $S^{flow} = 0.89$, $S^{activity} = 0.92$, and $S^{trips} = 0.90$. In this case, the simulation captures the major structure of patterns of life with minor calibration gaps.
Austin, on the other hand, has a low score $S^{trips} = 0.66$. 
This structural mismatch suggests poor behavioral fitting of the simulation and indicates that the simulation's parameters need to be calibrated to produce meaningful results.

In Cleveland, the standard set of parameters yields good similarity results: $S^{flow} = 0.88$, $S^{activity} = 0.91$, and $S^{trips} = 0.89$.
However, fine-tuning the parameters increases the descriptive power of the simulation, raising the similarity scores to $S^{flow} = 0.93$, $S^{activity} = 0.98$, and $S^{trips} = 0.96$.
Across metropolitan areas, the fact that most cities score above \(0.80\) indicates that patterns of life are broadly shared and can be reproduced without expensive city-by-city tuning.
However, targeted calibration yields additional, sometimes substantial, gains in descriptive performance when needed.

\begin{table}[t]
\centering
\caption{\textbf{Model with standard parameters captures patterns of life in U.S. cities.}
The similarity scores comparing simulation results with NHTS data show that most cities exhibit good agreement, with scores above \(0.80\).
All simulations were run using the same set of standard parameters, without city-specific tuning.
For Cleveland (adjusted), we also report the similarity scores after calibrating the model to better match empirical patterns, demonstrating that further improvements can be achieved through targeted calibration.}
\label{tab:sim-nhts-similarity-2dp}
\begin{tabular}{l
                S[table-format=1.2]
                S[table-format=1.2]
                S[table-format=1.2]}
\toprule
\textbf{City} & {\(S^{\mathrm{flow}}\)} & {\(S^{\mathrm{activity}}\)} & {\(S^{\mathrm{trips}}\)} \\
\midrule
12420--Austin              & 0.81 & 0.84 & 0.66 \\
17460--Cleveland (adjusted)           & 0.93 & 0.98 & 0.96 \\
17460--Cleveland (standard)    & 0.88 & 0.91 & 0.89 \\
18140--Columbus            & 0.88 & 0.88 & 0.94 \\
25540--Hartford            & 0.82 & 0.82 & 0.73 \\
33340--Milwaukee           & 0.83 & 0.85 & 0.73 \\
33460--Minneapolis         & 0.89 & 0.92 & 0.90 \\
35380--NewOrleans          & 0.86 & 0.88 & 0.84 \\
40140--Riverside           & 0.88 & 0.92 & 0.92 \\
\bottomrule
\end{tabular}
\end{table}

\section{\label{secDiscussion}Discussion}

One of the main challenges in the study of urban systems is representing the heterogeneous interactions between humans and city infrastructure.
These interactions can be modeled using agent-based models, where social interactions and responses to changes in the urban environment can be
effectively captured through the microscopic design of interaction rules.
However, the lack of universality and scalability of ABMs poses a significant challenge to the study of complex systems and the analysis of emergent collective behavior in large urban systems, as ABMs are computationally expensive.

In this work, we introduced a large-scale ABM to simulate patterns of life.
The empirical definition of ``patterns of life'' is an ill-posed problem without a universally accepted formulation.
Following the principles of Pattern-Oriented Modeling \cite{grimm2005pattern}, in which the validity of a model lies in its ability to reproduce a set of empirical patterns, we identified three main real-world patterns (extracted from the $2017$ U.S. National Household Travel Survey) to evaluate the performance of our model:
(1) the frequency of trips to activity destinations,
(2) the flow diagram of trips between origin–destination pairs, and
(3) the distribution of the number of trips per agent.
These patterns, further supported by the rank distribution of activity chains (see Supplementary Fig. 4) \cite{reia2025function, ectors2019exploratory}, suggest that the simulation we propose here is able to capture basic patterns of life.

The accuracy of the simulation depends not only on the design of the interaction rules of the agent-based model but also on the quality of the input data.
To generate realistic behavior, we designed a complete framework including infrastructure extraction, synthetic population generation, and social network creation.
The infrastructure extraction comprises data collection from OpenStreetMap, with subsequent building classification into the categories of the model (residential, workplace, recreational place, restaurant, school).
The synthetic population generation is based on Census tract demographic and employment data, which allows us to determine not only the number of men and women of each census tract, but also the population composition in terms of the agent types explicitly defined in our model: workers, students, and homemakers.
The social network is based on the spatial distance between agents, with agents closer two each other being more likely to be friends.
In this sense, our model can be used to test and explore the consequences of changes in infrastructure or population shifts, and also be improved with more accurate representations of city infrastructure and population.

Empirical patterns from NHTS were used to verify the results of our model.
Given the correlation between agent types (as workers) and fixed behaviors (trips to workplaces), we examined the similarities and differences in population composition between two empirical data sources, (i) Census sociodemographic data, which is used to generate the synthetic agents in our simulation, and (ii)the National Household Travel Survey, which is used to validate the model's results.
Although discrepancies in population composition between these two sources limit the cases in which the NHTS can be used for validation, we identified a metropolitan area (Cleveland) where the differences are small enough to enable a reasonable comparison.

The validation metrics we introduce here help us assess the transferability and universality of the simulation, but this notion of universality can capture non-linear collective behavior only if we can compute these similarities at the metropolitan scale.
Our simulation addresses the scalability challenge by leveraging multiple processing cores through the Message Passing Interface (MPI) protocol, implemented via Repast4Py \cite{collier2022distributed}.
Determining the runtime as a function of the number of agents is non-trivial, as it also depends on the complexity of the urban infrastructure and the distribution of nodes across the computing cluster.
This framework allows us to optimize the agent-based model and scale it to more than 20 million agents, as in the case of New York City (see Supplementary Table 6, Supplementary Fig. 11).

In summary, SimPOL functions as a ``behavioral engine'' that links micro-level decision rules with macro-level patterns of life outcomes. 
By combining individual routines, stochastic decision-making, and social interaction, the model generates complex city-scale dynamics such as temporal rhythms of activity and heterogeneous life patterns across demographic groups. 
These emergent properties highlight how accessibility and urban form shape everyday mobility. 
Beyond validation with existing survey data, the framework provides a flexible testbed for exploring scenarios of infrastructural change, demographic shifts, and policy interventions. 
In doing so, SimPOL offers not only a methodological contribution to large-scale activity-based modeling but also a practical tool for supporting more resilient urban mobility systems.


\section{Methods}\label{secMethods}

The model we propose creates a virtual representation of a city, consisting of two components, population and infrastructure.
Given the role of population composition in driving behavior and the role of infrastructure layout in constraining mobility patterns, accurately representing population composition and infrastructure characteristics leads to more realistic patterns of life.
The introduced framework for collecting and processing empirical data is based on Census and OpenStreetMap datasets and determines population demographics and infrastructure composition (Supplementary Fig. 1) at the metropolitan area level.

The population generation process begins with identifying the census tract delineations \cite{uscb2020tigerTracts} of a metropolitan area \cite{uscb2025metroMicro}.
According to \cite{census2020_tracts_2018}, census tracts are ``relatively permanent small-area geographic divisions of a county or statistically equivalent entity defined for the tabulation and presentation of data from the decennial census and selected other statistical programs.''
Census tracts have been widely used in health \cite{freeman2011association} and sociodemographic \cite{dai2017vape} studies, as they allow for a high spatial resolution representation of the processes taking place in cities.
For each census tract, we extract the population total and the population composition, which gives the fraction of males, females, age distribution, and workers by age in the area.
This information allows us to determine the number of workers, students, and homemakers in each tract, which are the agent types considered in our simulation. 

Once the building infrastructure of the metropolitan area is extracted from OpenStreetMap (OSM) using OSMNx \cite{boeing2017osmnx}, census tract delineations are used to identify the set of buildings within each tract.
The population of each tract is then allocated to the available residential buildings, with the number of residents per building assigned proportionally to its area.
Buildings are classified as residential or non-residential following the method introduced in \cite{f2024openstreetmap}, which uses OSM metadata to classify buildings.
Additional metadata associated with non-residential buildings are used to further categorize them as workplaces, schools, restaurants, or recreational places, which are the categories explicitly considered in our model.



\subsection{Building classification}
We extend the framework proposed in~\cite{f2024openstreetmap} to classify the building infrastructure extracted from OpenStreetMap using OSMnx. 
The building's classification as residential or non-residential is determined based on a combination of building footprints, OSM tags, and other spatial data, such as overlapping land-use polygons and points of interest.
OSM tags are also used to classify non-residential buildings further as workplaces, schools, restaurants, or recreational places, according to the simulation's specific activities.
The classification of schools, restaurants, and recreational places is based on a predefined mapping table in which tags are directly mapped to categories.
For example, a building tagged with ``building:school'' is classified as a school (see Supplementary Table 4 for details).
Finally, all buildings initially identified as non-residential but not reclassified into one of the previous three categories are assigned to the workplace category.

\subsection{Synthetic population generation}

We introduce a synthetic population generation method that integrates demographic data from U.S. Census \cite{USCensus2020DP1}, building infrastructure information from OpenStreetMap \cite{OpenStreetMap}, and Census tract-level employment statistics \cite{USCensusACS2017S2301}. 
The process we have used here makes sure that agents in the model were not only demographically representative of real-world populations but also spatially distributed within residential buildings within each Census tract, which are then used to make workplace and school assignments (see Supplementary Fig. 1).

The first stage of our method identifies the set of Census tracts that belong to a given metropolitan area, along with the buildings located within each tract. 
For each tract, we retrieve the set of residential buildings and then extract detailed age- and gender-specific population counts to spatially assign the population to residential buildings in the same tract.
Each residential building is weighted by its two-dimensional footprint area, and individuals are probabilistically assigned to buildings within their tract according to these weights, so that larger buildings have larger populations.
By doing this, we are able to generate a synthetic population of agents with assigned home locations that has real-world demographic information at the Census tract level.

To further characterize agents as workers, students or homemakers, we use employment data from the American Community Survey \cite{USCensusACS2017S2301}, which provides employment rates based on gender and age at the census tract level.
Based on these rates, agents aged 16 and above are probabilistically designated as workers or homemakers, while children under 16 are labeled as students.
The third category of agents in our simulation is homemakers, composed of agents not labeled as students or workers.
In the real world, this category captures individuals who are unemployed or retired.

For spatial realism in daily activity patterns, workers and students are designated to workplaces and schools.
To determine workplaces and schools of agents, we use spatial coordinates of these buildings to construct a KDTree~\cite{de2008computational} to efficiently match agents to buildings.
Workers are assigned to workplaces within a $15$ km radius of their homes when possible, or otherwise to the nearest workplace outside that radius. 
This distance threshold reflects the average trip length for most activity types and accounts for about $80\%$ of all trips in our dataset (see Supplementary Fig. 16). 
Students are always assigned to the school building nearest to their home.

The final synthetic population includes each agent’s identifier (id), gender, age, agent type (worker, student, or homemaker), and the spatial identifiers of their home, workplace, and/or school. 
This information is stored in a structured file and serves as the input for our agent-based simulation. 
The integration of multiple data sources, along with spatially grounded assignment processes, ensures that the synthetic population reflects both the demographic structure and spatial dynamics of the urban environment under study.

\subsection{Social network}

The social network is modeled as an undirected graph $G = (V, E)$, where the set of nodes $V = \{1, \dots, n\}$ represents individual agents, and the set of edges $E \subseteq V \times V$ denotes mutual friendships. 
To generate the network, we assume that agents who live and work in the same buildings are more likely to be connected. 
To capture this spatial dependency, we use an adapted version of a soft random geometric graph~\cite{penrose2016connectivity} (soft RGG), in which the probability of an social connection between two agents depend on the Euclidean distance between their home or work spatial locations. 

A soft RGG is defined by placing $n$ nodes at random positions $\{x_i\}_{i=1}^n$ in a 2-dimensional space, sampled from a distribution $P$. Edges are formed independently with probability
\begin{equation}
\mathbb{P}[(x_i, x_j) \in E] = f(\|x_i - x_j\|),
\end{equation}
where $f : [0, \infty) \to [0,1]$ is a non-increasing connection function. Unlike the hard RGG~\cite{penrose2003random}, which connects nodes within a fixed radius, the soft RGG allows for probabilistic connectivity that decays with distance. Here, we adopt a Gaussian decay function:
\begin{equation}
\mathbb{P}[(x_i, x_j) \in E] = \exp\left(-\frac{\|x_i - x_j\|^2}{2 r^2}\right),
\end{equation}
where $r > 0$ is a scale parameter that controls the connectivity range.

Here, we create social connections between pairs of agents based on their home and work locations. 
Specifically, we construct two separate graphs: one representing home locations ($G_h$) and the other representing workplace locations ($G_w$). 
In $G_h$, we represent social connections between agents living in nearby locations. 
To avoid unrealistically high connectivity between agents in the same building, we introduce a small spatial perturbation by adding 30 meters of random jitter to each location. 
Then, the home graph $G_h$ is generated using the soft RGG model described above. 
We apply the same procedure to generate $G_w$, which is the social network of worker agents considering the distance between their work locations. 
The resulting social network $G$ is given by merging the edge sets of $G_h$ and $G_w$.


The generated social network exhibits a high clustering coefficient~\cite{watts1998collective}, reflecting the tendency for neighbors of a node to also be connected, thereby forming triangles. This property is characteristic of empirical social networks~\cite{newman2003social}. 
Furthermore, we define the average network degree (i.e., the average number of friends) as 30 because it lies between the sympathy group and good friends in Dunbar's theory~\cite{dunbar1993coevolution}.
To achieve this, we control the scale parameter $r$ in the connection probability function, which determines the spatial reach of potential friendships. Since we construct two separate networks based on home and workplace locations, we assign separate parameters: $r_h$ for the home graph $G_h$ and $r_w$ for the work graph $G_w$. For simplicity and to reflect the broader spatial spread of residential ties compared to workplace interactions, we set $r_h = 3 r_w$. The parameters used to generate the networks, along with their features, are shown in Supplementary Table 5. 

Once the social network is generated, it is used as input to our simulation and remains static over time. The assumption of a stable friendship network over the course of the simulation (which is run on daily time scales) is supported by empirical evidence showing that friendship networks are highly stable on yearly time scales \cite{roy2022turnoverclosefriendshipsage}.

\subsection{Agent-Infrastructure Mapping}

Empirical studies show that human travel patterns are highly predictable~\cite{science_limits}, and that individuals tend to visit a fixed number of different locations at any point in time~\cite{alessandretti2018evidence}.
The exploratory nature of human behavior reveals that the number of unique locations visited by individuals growth smoothly over time~\cite{pappalardo2015returners}, with people visiting from $30$ to $60$ unique locations on a period of $2$ months~\cite{alessandretti2018evidence}.
Based on that, and to optimize the computational performance of the simulation by constraining the search space of the agents, each agent in the simulation is pre-assigned a set of $10$ restaurants, $10$ recreational places, and $20$ errand places that can be visited during the simulation.
Errands places are composed of residential and workplaces other than the agent's own home and workplace, capturing non-explicit trips to all the other categories described in the NHTS.
We set these numbers to match the number of unique places individuals are reported to visit in a time frame of $10$ to $100$ days in \cite{alessandretti2018evidence}, and can be easily modified to better recover shorter or longer time frames.
Once again, using the home location as a reference point, agents randomly select restaurants, recreational places and errand places within a $15km$ radius.

\subsection{SimPOL Agent‐Based Model Description}

The SimPOL framework models short-term urban mobility as the interplay between mandatory (fixed) activities, flexible activities driven by evolving needs, spatial constraints given by the spatial distribution of building types, and social influences. 
At its core, the simulation captures how individuals balance mandatory activities (work or school) with flexible, needs-driven activities (food, errands, and recreation) in different urban settings.

\subsubsection{Initialization and Environment}

The SimPOL is designed in Python around the agent-based framework Repast4py.
The model begins by constructing a shared spatial environment defined by the centroids of classified buildings (e.g., residential, workplaces, restaurants, recreational, and errands locations). 
This environment is distributed across MPI processes to enable large-scale parallel execution. 
A global scheduler controls the temporal evolution of the system, ensuring synchronization across all ranks. 
Agents are instantiated with demographic attributes, fixed home and work/school locations, and individualized lists of potential destinations (viz. 10 restaurants, 10 recreational places, 20 errand places) for flexible activities.

\subsubsection{Agent State and Needs.}

Each agent can be an object of the Worker, Student, or Homemaker classes, and is instantiated with:

\begin{itemize}
  \item \textit{Static locations} such as home and workplace/school;
  \item \textit{Dynamic needs} for food, recreation, and errands, which accumulate over time according to linear decay functions;
  \item \textit{A daily activity buffer} containing planned destinations, travel times, and dwell durations;
  \item \textit{Mobility parameters} such as transportation mode and expected trip durations.
\end{itemize}

These elements define the internal state of each agent, which evolves continuously through assessment of needs, opportunities, and constraints.

\subsubsection{Daily Scheduling.}

Agents operate on a daily rhythm. Each morning, they ``wake up'' at a stochastic time and determine whether work or school will occupy the first part of the day. 
After mandatory activities are set, agents evaluate their unmet needs. 
If a need exceeds its satisfaction threshold, the agent schedules a flexible activity. 
Destination selection follows a probabilistic mechanism (following a Zipf probability distribution), where most liked locations are more likely to be selected, introducing heterogeneity and reflecting the empirical fact that certain locations disproportionately attract more visits.

\subsubsection{Movement and Interaction.}

Movement unfolds tick by tick, where each tick represents 5 minutes:

\begin{enumerate}
  \item Agents use travel time as they move between locations;
  \item Upon arrival, they update their state and log their location;
  \item Social reinforcement occurs: agents co-located with friends in the same recreational place experience increased social satisfaction, which then modulates when agents go to recreational places again;
  \item If an agent crosses a spatial partition, its state is seamlessly migrated to the appropriate MPI rank thanks to Repast4Py framework.
\end{enumerate}

This stepwise execution enables both micro-scale trajectories (individual trips) and macro-scale emergent patterns, which can be verified using flow data as the NHTS.




\section*{Data Availability}

All empirical data used to parameterize and validate the model are drawn from publicly available sources.  
Daily travel behavior is taken from the 2017 U.S. National Household Travel Survey (NHTS) \cite{NHTS2017}.  
Demographic and employment information at the census-tract level are obtained from the 2020 Decennial Census \cite{USCensus2020DP1} and the 2013--2017 American Community Survey 5-year estimates (Table S2301) \cite{USCensusACS2017S2301}.  
Census tract and metropolitan area boundaries are based on U.S. Census Bureau TIGER/Line products and tract delineations \cite{uscb2020tigerTracts,uscb2025metroMicro,census2020_tracts_2018}.  
Building footprints and land-use attributes are derived from OpenStreetMap \cite{OpenStreetMap} and processed using OSMnx \cite{boeing2017osmnx} and the building-classification framework of \cite{f2024openstreetmap}.

All of these raw datasets can be downloaded directly from the original providers; no proprietary data are used.
Derived datasets generated in this study (including classified building layers, synthetic populations, and aggregated simulation outputs) are available from the corresponding author upon reasonable request.



\section*{Code Availability}

Computer codes will be made available upon reasonable request.

\section*{Funding}

This work was supported by the Intelligence Advanced Research Projects Activity (IARPA) via Department of Interior/Interior Business Center (DOI/IBC) contract number 140D0419C0050. The U.S. Government is authorized to reproduce and distribute reprints for Governmental purposes notwithstanding any copyright annotation thereon. Disclaimer: The views and conclusions contained herein are those of the authors and should not be interpreted as necessarily representing the official policies or endorsements, either expressed or implied, of IARPA, DOI/IBC, or the U.S. Government.











\section*{Author Contributions}

S.M.R., T.A., H.K., and D.P. conceived the study, with T.A., H.K., and D.P. acquiring funding; D.P. provided resources and administered the project. S.M.R., H.F.A., and S.R. curated the data. S.M.R. and H.F.A. performed the formal analysis. S.M.R., H.F.A., S.R., T.A., H.K., and D.P. conducted the investigation and contributed to methodology. S.M.R. carried out validation and prepared the visualizations. S.M.R., H.F.A., T.A., H.K., and D.P. wrote the original draft. All authors reviewed and edited the manuscript.

\section*{Competing Interests}
The authors declare no competing interests.

\newpage

\bibliography{refs}

\end{document}